\documentclass{iopart}
\usepackage{bm}
\usepackage{graphicx}
\usepackage{subfigure}
\usepackage{multirow}
\begin{document}
%
%
%
%
\title{Bond percolation on a class of correlated and clustered random graphs}
\author{A Allard, L H\'ebert-Dufresne, P-A No\"el, V Marceau and \mbox{L J Dub\'e}}
\address{D\'epartement de physique, de g\'enie physique, et d'optique, Universit\'e Laval, Qu\'ebec (Qc), Canada G1V 0A6}
\ead{antoine.allard.1@ulaval.ca}
\begin{abstract}
 We introduce a formalism for computing bond percolation properties of a class of correlated and clustered random graphs. This class of graphs is a generalization of the Configuration Model where nodes of different types are connected via different types of \emph{hyperedges}, edges that can link more than 2 nodes. We argue that the multitype approach coupled with the use of clustered hyperedges can reproduce a wide spectrum of complex patterns, and thus enhances our capability to model real complex networks. As an illustration of this claim, we use our formalism to highlight unusual behaviors of the size and composition of the components (small and giant) in a synthetic, albeit realistic, social network.
\end{abstract}
%
%
%
%
%
\section{Introduction}
%
Bond percolation is the study of the size distribution of components in graphs whose edges exist with a given probability. For its theoretical appeal and its varied applications in many contexts, mathematical modelling of bond percolation on random graphs has recently received substantial attention (see \cite{Dorogovtsev08_RevModPhys,newman10_Networks}, and references therein). Within the Configuration Model (CM) paradigm \cite{Molloy95_RandomStructAlg,Molloy98_CombinatoricsProbabComput}, many exact results can be obtained using probability generating functions (PGF) \cite{Newman01_PhysRevE}. This analytic tractability however comes at the price of simplifying assumptions on the structure of the graphs.

We introduce a generalization of the CM that encompasses many of the previous improvements published to this day \cite{Newman01_PhysRevE,Allard09_PhysRevE,Ghoshal09_PhysRevE,Gleeson09_PhysRevE,Karrer10_PhysRevE,Leicht09_arXiv,Miller09_PhysRevE,Meyers06_JTheorBiol,Newman02_PhysRevE,Newman02_PhysRevLett,Newman03a_PhysRevE,Newman03b_PhysRevE,Newman09_PhysRevLett,Serrano06_PhysRevLett,Shi07_PhysicaA,Zlatic12_EPL}, and brings this class of models closer to the behavior of real complex networks. By combining the multitype approach of \cite{Allard09_PhysRevE}, the analytical method of \cite{Allard12_EPL} and the one-mode projection of \cite{Newman03b_PhysRevE}, we argue that our model is able to reproduce a wide range of complex patterns found in real networks.

On the one hand, the multitype approach allows to explicitly prescribe how nodes are connected to one another in a very detailed fashion. By assigning types to nodes -- in other words by knowing who is who, and therefore who is connected to whom -- several mixing patterns (e.g., assortativity, degree correlation, node segregation), as well as heterogeneous bond occupation probabilities (e.g., partial and/or uneven directionnality of edges) can be reproduced. On the other hand, the use of the one-mode projection, coupled with the multitype approach, allows the inclusion of clustering through a myriad of nontrivial motifs, i.e. recurrent, significant patterns of interconnections \cite{Milo02_Science}.

This paper is organized as follows. In \sref{sec:themodel}, we introduce the generalization of the CM that explicitly includes various correlations and clustering. We then develop the analytical framework to obtain the bond percolation properties of this graph ensemble in \sref{sec:math_form}. In \sref{sec:validation}, we validate our formalism --- and also illustrate the versatility of our approach --- by comparing its predictions with simulation results on a synthetic, but realistic social network. In \sref{sec:special_cases}, we show that many percolation models published in the litterature are special cases of our model. We also highlight how our approach can be useful for studying interdependent or coupled networks \cite{Leicht09_arXiv,Buldyrev10_Nature,Huang11_PhysRevE,Marceau11_PhysRevE},  and for studying the \textit{weak and strong clustering regimes} \cite{Serrano06a_PhysRevE,Serrano06b_PhysRevE}. We conclude in \sref{sec:conclusion} and present in 2 Appendices some relevant aspects of the analysis and simulations. \ref{sec:general_method} details how a recent method, to analytically compute the distribution of the composition of components for any small arbitrary graphs \cite{Allard12_EPL}, can be used in our formalism. \ref{sec:num_sim_details} gives further details on the numerical simulations.
%
%
%
%
%
\section{Correlated and clustered graph ensemble} \label{sec:themodel}
%
We introduce a general class of correlated and clustered random graphs. To preserve the analytical tractability of the CM, we first consider \textit{unclustered multitype bipartite graphs} that are locally tree-like in the large system size limit. Clustering is then incorporated through a projection, analogous to the one-mode projection of \cite{Newman03b_PhysRevE}.
%
%
%
\subsection{Unclustered multitype bipartite graph ensemble}
\begin{figure}[tb]
 \centering
 \subfigure[]{\label{fig:allard_fig_1a} \includegraphics[width = 0.4\textwidth]{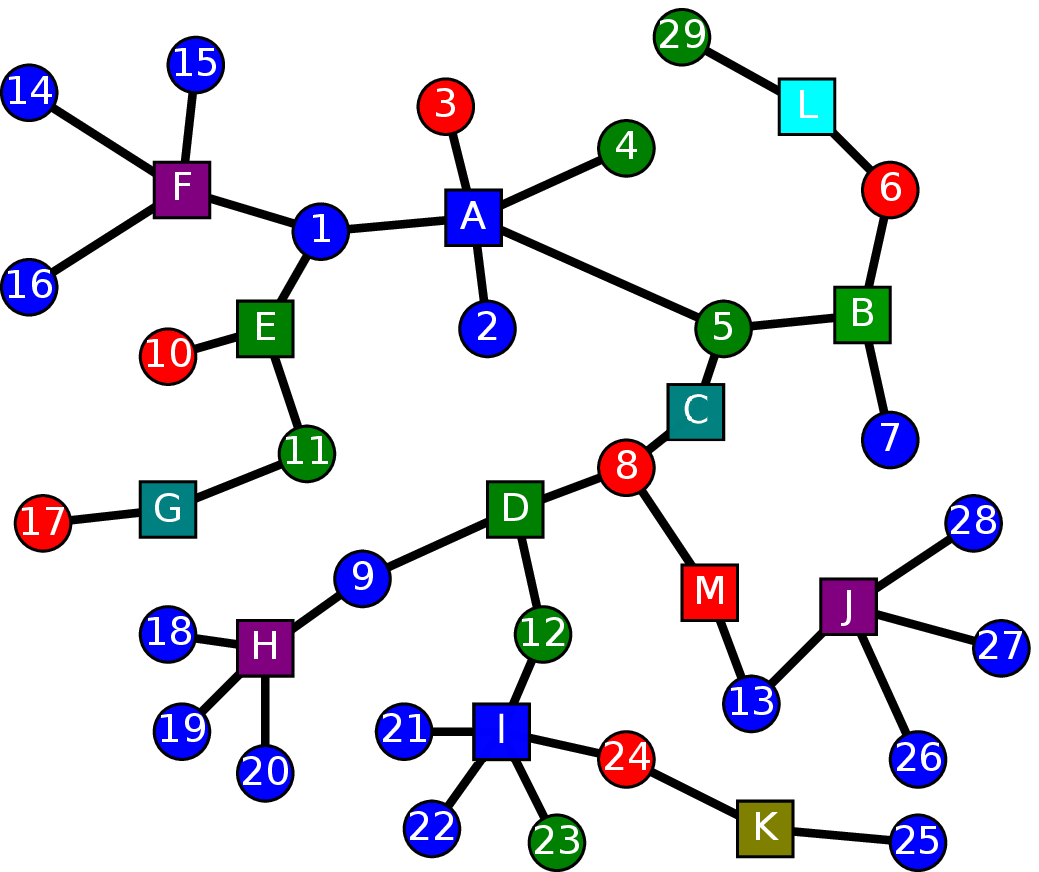}} \hspace{0.15\textwidth}
 \subfigure[]{\label{fig:allard_fig_1b} \includegraphics[width = 0.4\textwidth]{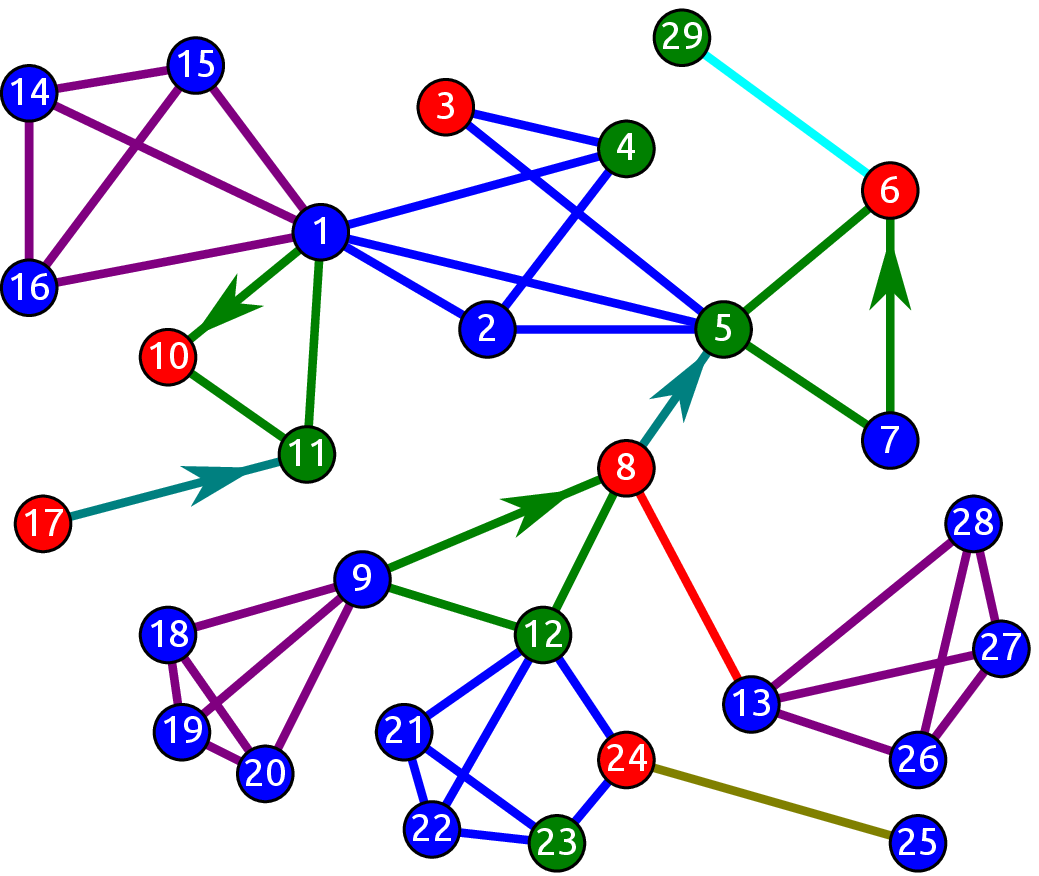}}
  \caption{\label{fig:allard_fig_1}(colour online) Illustration of the projection process introduced in \sref{sec:themodel}. (a) In unclustered multitype bipartite graphs, nodes (29 circles) belong to different types (colours, $M=3$) and are linked exclusively to groups (13 squares), which are distinguished through types as well (colours, $\Lambda=7$). (b) In clustered multitype graphs, nodes linked to a same group in the underlying unclustered multitype bipartite graph are linked to one another through a motif whose nature and structure are specified by the corresponding group type. Labels have been added to nodes and groups for the sake of comparison between (a) and (b) and are not part of the model.}
\end{figure}

%
We call unclustered multitype bipartite graphs a multitype generalization of the bipartite CM \cite{Newman01_PhysRevE}. These graphs are composed of $M$ types of ``regular nodes'' and $\Lambda$ types of ``group nodes'' (hereafter referred to as nodes and groups, respectively). Edges only exist between regular nodes and group nodes. In these graphs, a fraction $w_i$ of nodes are of type $i$, and any given type-$i$ node is connected to $k_\mu$ type-$\mu$ groups (for each $\mu = 1, \ldots, \Lambda$) with a probability $P_i(k_1,\ldots,k_\Lambda) \equiv P_i(\bm{k})$. Likewise, a randomly chosen type-$\nu$ group is connected to $n_j$ type-$j$ nodes (for each $j = 1, \ldots, M$) with a probability $R_\nu(n_1,\ldots,n_M) \equiv R_\nu(\bm{n})$. In other words, $R_\nu(\bm{n})$ is the distribution of the composition of type-$\nu$ groups. \Fref{fig:allard_fig_1a} gives an example of such graphs. To lighten the notation, it should now be understood that any free latin (resp. greek) index can take any values in $\{1, \ldots, M\}$ (resp. $\{1, \ldots, \Lambda\}$), except if otherwise mentionned.

In the large system size limit, $w_i$, $P_i(\bm{k})$ and $R_\nu(\bm{n})$ fully define a graph ensemble which is totally random in all other respects (stubs are matched randomly). All finite components therefore have a tree-like structure in this limit (the probability of a closed path goes as the inverse of the size of the graph). These quantities are however not independent. To guarantee the consistency of the graph ensemble, they must, for all applicable combinations of $i$, $j$ and $\nu$, satisfy
\begin{equation} \label{eq:consistency_condition}
 \frac{w_i \langle k_\nu \rangle_{P_i}}{\langle n_i \rangle_{R_\nu}} = \frac{w_j \langle k_\nu \rangle_{P_j}}{\langle n_j \rangle_{R_\nu}} \ ,
\end{equation}
where $\langle a \rangle_B$ denotes the mean value of $a$ with respect to the distribution $B$. Simply stated, \eref{eq:consistency_condition} asks $w_i$, $P_i(\bm{k})$ and $R_\nu(\bm{n})$ to be chosen such that each node type ``forces'' the same number of type-$\nu$ groups in the unclustered multitype bipartite graph ensemble.
%
%
%
\subsection{Clustered multitype graph ensemble}
%
A clustered graph ensemble is obtained from the unclustered multitype bipartite graph ensemble by means of a projection similar to the one-mode projection of \cite{Newman03b_PhysRevE}. This projection is achieved by replacing the group nodes in the unclustered multitype bipartite graphs by motifs involving the nodes that were linked to a same group. The nature, either quenched (fixed) or annealed (random), and the structure of these motifs is prescribed by the corresponding group type. The resulting graphs then consist of different motifs embedded in a tree-like backbone.

\Fref{fig:allard_fig_1b} illustrates a resulting clustered multitype graph where every group is replaced by a multitype quenched motif. For instance, type-\textit{green} groups (B, D and E) are replaced by a \textit{triangle} composed of one node of each of the $M=3$ possible types, and whose edges are undirected except for the one between the type-\textit{blue} node and type-\textit{red} one that is directed. Single edges --- directed (C and G) or undirected (K, L, and M) --- can also correspond to motifs simply composed of two nodes. The type of each of the two nodes and the direction of the edge is prescribed by the type of the group. An example of the use of annealed motifs, where edges exist with a given probablity rather than being \textit{a priori} fixed, is given in \sref{sec:validation}.

Bond percolation can exactly be solved for the CM and its numerous variants because graphs in these ensembles have an underlying tree-like structure. Thus to take advantage of the tree-like backbone of the clustered graph ensemble, the outcome of bond percolation must be solved beforehand for each motif appearing in the graph ensemble. This solution is encoded in $Q_{i\nu}(\bm{l}|\bm{n})$ giving the probability that $\bm{l}$ nodes (i.e., $l_j$ type-$j$ nodes, for all $j$) will eventually be reached from an initial type-$i$ node by following existing edges in a type-$\nu$ motif of size $\bm{n}$. In other words, this distribution prescribes the number of nodes (and their type) from which a given motif can be left while navigating on a graph of the clustered ensemble. It therefore ``restores'' the tree-like structure of the unclustered multitype bipartite graphs while retaining the effect of the clustered motifs. It is this correspondence that allows the derivation of a PGF-formalism which exactly solves the bond percolation properties of the clustered multitype graph ensemble.

In principle, a wide variety of motifs can be incorporated in our model; this variety is only limited by our ability to solve the bond percolation outcome on these motifs. Motifs can be chosen to reproduce recurring patterns of interactions found in real complex networks \cite{Milo02_Science}, to account for local clustering in realistic synthetic networks (see \sref{sec:validation}), or for theoretical investigations (see \sref{sec:weak_and_strong}). Following the results of \cite{Allard12_EPL}, we give in \ref{sec:general_method} a general method to calculate $Q_{i\nu}(\bm{l}|\bm{n})$ for most, if not all, imaginable motifs of reasonable size (the limits of the method are discussed in \cite{Allard12_EPL}). This method can handle quenched (fixed structure) or annealed (random structure) motifs in which edges may be directed or not. Also, nodes may belong to types which permits to model (dis-)assortative mixing and heterogeneous bond percolation \cite{Allard09_PhysRevE,Allard12_EPL}.
%
%
%
%
%
\section{\label{sec:math_form}Bond Percolation Properties}
%
We now introduce a PGF-based mathematical formalism to calculate the percolation properties of the correlated and clustered graph ensemble defined in the last section. Since PGF-based percolation formalisms have become fairly standard, the unfamiliar reader should consult recent reviews on complex network modeling (see for example \cite{Newman03_SIAMRev} and references therein) for further details.

We first define $\theta_{i\nu}(\bm{x})$ as the function generating the distributions $\{Q_{i\nu}(\bm{l}|\bm{n})\}$ of the outcome of bond percolation, from an initial type-$i$ node, on the motifs corresponding to type-$\nu$ groups. As type-$\nu$ groups may not all have the same composition (e.g., household size distribution in social networks), $\theta_{i\nu}(\bm{x})$ is calculated according to
\begin{equation} \label{eq:theta_general}
 \theta_{i\nu}(\bm{x}) = \sum_{\bm{n}} \frac{n_i R_{\nu}(\bm{n})}{\langle n_i \rangle_{R_\nu}} \left[ \sum_{\bm{l}=\bm{\delta_i}}^{\bm{n}} Q_{i\nu}(\bm{l}|\bm{n}) \prod_{j} x_{\nu j}^{l_j-\delta_{ij}} \right] \ ,
\end{equation}
with $\bm{\delta_i} \equiv (\delta_{i1},\ldots,\delta_{iM})$ where $\delta_{il}$ is Kronecker's delta. In \eref{eq:theta_general}, we average over $\frac{n_i R_{\nu}(\bm{n})}{\langle n_i \rangle_{R_\nu}}$ instead of over $R_{\nu}(\bm{n})$ to account for the fact that groups containing $n_i$ type-$i$ nodes are $n_i$ times more likely to be reached from any type-$i$ node than groups containing only one type-$i$ node. Although \eref{eq:theta_general} is not explicitly labelled in this respect, more than one motifs may be associated with a given group type. In such case, the distribution $R_\nu(\bm{n})$ gives the probability of occurence of each motif, and the left-hand sum in \eref{eq:theta_general} is taken over each possible motif for which a distinct distribution $Q_{i\nu}(\bm{l}|\bm{n})$ is obtained with the method outlined in \ref{sec:general_method}.

The function $\theta_{i\nu}(\bm{x})$ is the mathematical implementation of the correspondence between the unclustered and clustered graph ensembles discussed at the end of the last section. By generating the distribution of $\nu \rightarrow j$ edges (i.e., stemming from a type-$\nu$ group and leading to a type-$j$ node) reached by a type-$i$ node, one can then navigate on a unclustered multitype bipartite graph as if one were on a clustered multitype graph.

We define $g_i(\bm{x})$ as the PGF generating the distribution of the number of $\nu \rightarrow j$ edges emerging from a type-$i$ node (i.e., emerging from the groups a type-$i$ node is connected to)
\begin{equation}
 g_i(\bm{x}) = \sum_{\bm{k}} P_i(\bm{k}) \prod_\nu \big[\theta_{i\nu}(\bm{x})\big]^{k_\nu} \ .
\end{equation}
It is also convenient to define a PGF that generates the distribution of the number of $\nu \rightarrow j$ edges emerging from a type-$i$ node which has itself been reached via a $\mu \rightarrow i$ edge
\begin{equation} \label{eq:fmui}
 f_{\mu i}(\bm{x}) = \sum_{\bm{k}} \frac{k_\mu P_i(\bm{k})}{\langle k_\mu \rangle_{P_i}} \prod_{\nu} \big[\theta_{i\nu}(\bm{x})\big]^{k_\nu - \delta_{\mu\nu}} \ .
\end{equation}
The averaging term used in \eref{eq:fmui} is motivated by the same argument as the one in \eref{eq:theta_general}. With these two PGFs, we may now compute the percolation properties of clustered multitype graph ensemble.
%
%
%
\subsection{Phase transition}
%
As a class of random graphs, clustered multitype graphs undergo a phase transition corresponding to the emergence of an extensive connected ``giant'' component. To locate the phase transition, let us define $\xi_{\nu j}(s)$ as the average number of $\nu \rightarrow j$ edges at a distance $s$ from any node in any graphs of the ensemble. Due to the tree-like structure of the underlying unclustered multitype bipartite graph, each $\xi_{\nu j}(s)$ is a linear combination of all $\xi_{\nu j}(s\!-\!1)$ at distance $s\!-\!1$:
\begin{equation} \label{eq:phase_transition1}
 \xi_{\nu j}(s) = \sum_{\mu i} \left[ \frac{\partial f_{\mu i}(\bm{x})}{\partial x_{\nu j}}{} \right]_{\bm{x}=\bm{1}} \xi_{\mu i}(s\!-\!1)
\end{equation}
where
\begin{equation}
 \frac{\partial f_{\mu i}(\bm{1})}{\partial x_{\nu j}}{} = \sum_{\bm{k}} \frac{(k_\nu\!-\!\delta_{\mu\nu})k_\mu P_i(\bm{k})}{\langle k_\mu \rangle_{P_i}} \frac{\partial \theta_{i\nu}(\bm{1})}{\partial x_{\nu j}}
\end{equation}
is the average number of $\nu \rightarrow j$ edges emerging from a type-$i$ node that has been reached via a $\mu \rightarrow i$ edge. In vector notation, \eref{eq:phase_transition1} becomes
\begin{equation} \label{eq:phase_transition2}
 \bm{\xi}(s) = \mathbf{B}\, \bm{\xi}(s\!-\!1) \ .
\end{equation}
We see from \eref{eq:phase_transition2} that, in general, every $\xi_{\nu j}(s)$ vanishes with increasing $s$ if all eigenvalues of the $(M\Lambda) \times (M\Lambda)$ matrix $\mathbf{B}$ are below 1. Thus the phase transition happens when the largest eigenvalue of $\mathbf{B}$ reaches unity \footnote{We see from \eref{eq:phase_transition1} that $\mathbf{B}$ is a non-negative and, in general, irreducible matrix. Thus the Perron-Frobenius theorem \cite{Meyer00_MatrixAnalysis} ensures that the largest eigenvalue of $\mathbf{B}$ is simple, real and positive. Moreover, the associated eigenvector is the only nonnegative eigenvector of $\mathbf{B}$.}.
%
%
%
\subsection{Giant Component}
%
As there may be directed edges in the graphs (through the motifs), the giant component may have a ``bow-tie'' structure \cite{Newman01_PhysRevE,Allard09_PhysRevE}. This implies that the probability $\mathcal{P}$ of reaching the giant component may not be equal to its relative size $\mathcal{S}$. Both quantities must therefore be computed separately.

Let us define $a_{\mu i}$ as the probability that a $\mu \rightarrow i$ edge does not lead to the giant component. Because of the tree-like structure of finite components in the unclustered multitype bipartite graph, we see that $a_{\mu i}$ must satisfy the self-consistency relation
\begin{equation} \label{eq:gc_selfconsistent}
 a_{\mu i} = f_{\mu i}(\bm{a}) \ .
\end{equation}
That is, every edge reached from an edge that is not leading to the giant component must not lead to the giant component either. The probability that any type-$i$ node does lead to the giant component is therefore given by $\mathcal{P}_i \equiv 1-g_i(\bm{a})$, and, averaging over the node type distribution $\{w_i\}$, the probability $\mathcal{P}$ that a randomly chosen node leads to the giant component is
\begin{equation} \label{eq:P}
 \mathcal{P} = \sum_i w_i \mathcal{P}_i = 1 - \sum_{i} w_i g_i(\bm{a}) \ .
\end{equation}

To obtain the size of the giant component, we must calculate the probability that a given node cannot be reached from any node in the giant component. This is equivalent to computing the probability that this node does not lead to the giant component when edges are followed in the reverse direction \cite{Newman01_PhysRevE,Allard09_PhysRevE}. Edges in the underlying unclustered multitype bipartite graph being undirected, only $\theta_{\nu i}(\bm{x})$ needs to be modified. For instance, this can be achieved by using $p_{sr}$ instead of $p_{rs}$ in \eref{eq:iterative_erdos_renyi_1}. We denote this new PGF $\bar{\theta}_{\nu i}(\bm{x})$ and we will add a bar ($\bar{\ }$) over every PGF using $\bar{\theta}_{\nu i}(\bm{x})$ instead of $\theta_{\nu i}(\bm{x})$.

Following a similar approach as for computing $\mathcal{P}$, we define $\bar{a}_{\mu i}$ as the probability that a type-$i$ node cannot be reached from the giant component via a $\mu \rightarrow i$ edge. That is, $\bar{a}_{\mu i}$ is the probability that a neighbour of a type-$i$ node in a type-$\mu$ group is not part of the giant component. Self-consistency then requires for $\bar{a}_{\mu i}$ to satisfy
\begin{equation}
 \bar{a}_{\mu i} = \bar{f}_{\mu i}(\bm{\bar{a}}) \ .
\end{equation}
The probability that any given type-$i$ node is not part of the giant component is therefore $\bar{g}_i(\bm{\bar{a}})$. Considering that a fraction $w_i$ of the nodes are of type $i$, the fraction of the graph occupied by type-$i$ nodes in the giant component is
\begin{equation} \label{eq:Si}
 \mathcal{S}_i = w_i \big[ 1 - \bar{g}_i(\bm{\bar{a}}) \big] \ ,
\end{equation}
and the relative size of the giant component is
\begin{equation}
 \mathcal{S} = \sum_i \mathcal{S}_i = 1 - \sum_{i} w_i \bar{g}_i(\bm{\bar{a}}) \ .
\end{equation}
%
%
%
\subsection{Distribution of the composition of small components}
%
To calculate the distribution of the number of nodes of each type expected in small components, we define the PGF $A_{\mu i}(\bm{x})$ that generates the distribution of the number of edges of each type (i.e., $\nu \rightarrow j$ for all $\nu$ and $j$) that are \textit{ahead} of a $\mu \rightarrow i$ edge in small components. In the large system size limit, the small components have a tree-like structure and no finite-size effects are to be expected [i.e., the joint degree distribution $P_i(\bm{k})$ is constant]. We therefore expect $A_{\mu i}(\bm{x})$ to be invariant under translation on a small component; the distribution of the number of each edge type ahead, $A_{\mu i}(\bm{x})$, is independent of the position in a small component. This implies that $A_{\mu i}(\bm{x})$ must satisfy
\begin{equation} \label{eq:sc_translation_independence}
 A_{\mu i}(\bm{x}) = x_{\mu i} f_{\mu i}\big(\bm{A}(\bm{x})\big)
\end{equation}
where the extra $x_{\mu i}$ accounts for the $\mu \rightarrow i$ edge that has just been followed. This extra factor guarantees that a finite extent of the distribution generated by $A_{\mu i}(\bm{x})$ can be obtained in a finite number of iterations of \eref{eq:sc_translation_independence} starting with the initial conditions $A_{\mu i}(\bm{x})=1$. Replacing $x_{\nu i} = z_i$ for all $\nu$ in $A_{\mu i}(\bm{x})$ generates the distribution of the number of nodes of each type ahead a type-$i$ nodes reached from a type-$\mu$ group. Thus the composition of a small component reached from a type-$i$ node is generated by $z_i g_i\big(\bm{A}(\bm{z})\big)$; again the extra $z_i$ accounts for the initial type-$i$ node. Because any node is of type $i$ with probability $w_i$, the composition of a small component that is reached from a randomly chosen node is therefore generated by
\begin{equation} \label{eq:small_components_size_dist}
 K(\bm{z}) = \sum_i \frac{w_i z_i g_i\big(\bm{A}(\bm{z})\big)}{1-\mathcal{P}} \ ,
\end{equation}
where $1-\mathcal{P}$ ensures the normalization of $K(\bm{z})$. Note that $A_{\mu i}(\bm{1})$ is equal to the probability that a $\mu \rightarrow i$ edge leads to a finite (small) component, and is therefore equal to $a_{\mu i}$.

Solving \eref{eq:sc_translation_independence}--\eref{eq:small_components_size_dist} can however become tedious when dealing with large number of types of nodes and groups, or large groups. It is therefore worth noting that the first moments of the distribution generated by $K(\bm{z})$ can be calculated in a more direct manner. For instance, let us compute the average number $\langle s_i \rangle$ of type-$i$ nodes in small components. With
\begin{equation*}
 \langle s_i \rangle = \left. \frac{\partial K(\bm{z})}{\partial z_i}{} \right|_{\bm{z}=\bm{1}}
\end{equation*}
inserted in \eref{eq:small_components_size_dist}, replacing $x_{\mu i}$ with $z_i$, we get
\begin{equation} \label{eq:sc_average_1}
 \langle s_i \rangle = \frac{w_i (1\!-\!\mathcal{P}_i)}{1-\mathcal{P}}
            + \sum_{j \gamma r} \frac{w_j \langle k_\gamma\rangle_{P_j} a_{\gamma j}}{1-\mathcal{P}}
                \frac{\partial \theta_{j\gamma}(\bm{a})}{\partial x_{\gamma r}}{}
                \frac{\partial A_{\gamma r}(\bm{1})}{\partial z_i}{} \ ,
\end{equation}
where we have used \eref{eq:fmui}, \eref{eq:gc_selfconsistent} and the fact that $g_i(\bm{a}) = 1-\mathcal{P}_i$. In this last result, $\frac{\partial \theta_{j\gamma}(\bm{a})}{\partial x_{\gamma r}}{}$ is the average number of type-$r$ nodes that are accessible from a type-$j$ node in a type-$\gamma$ group in small components. Also, $\frac{\partial A_{\gamma r}(\bm{1})}{\partial z_i}{}$ is the average number of type-$i$ nodes ahead of a $\gamma \rightarrow r$ edge in small components. From \eref{eq:gc_selfconsistent}, we see that this last quantity is the solution of
\begin{equation} \label{eq:sc_average_2}
 \frac{\partial A_{\gamma r}(\bm{1})}{\partial z_i}{}
       = a_{\gamma r}\delta_{ir}
       + \sum_{\lambda s} \frac{\partial f_{\gamma r}(\bm{a})}{\partial \theta_{r\lambda}}{}
                          \frac{\partial \theta_{r\lambda}(\bm{a})}{\partial x_{\lambda s}}{}
                          \frac{\partial A_{\lambda s}(\bm{1})}{\partial z_i}{}
\end{equation}
where $\frac{\partial f_{\gamma r}(\bm{a})}{\partial \theta_{r\lambda}}{}$ is the average number of type-$\lambda$ groups to which a type-$r$ node reached via a type-$\gamma$ group is connected in small components. Thus by solving \eref{eq:gc_selfconsistent}--\eref{eq:P} and then \eref{eq:sc_average_1}--\eref{eq:sc_average_2}, it is possible to obtain quite easily the average number of nodes of each type in the small components. Equations for higher moments can be obtained in a similar manner and are straightforward to derive.
%
%
%
%
%
\section{Illustration and validation} \label{sec:validation}
%
To illustrate the versatility and the usefulness of our approach, we generated \textit{urban networks} \cite{Meyers05_JTheorBiol} and used our formalism to predict the outcome of an outbreak of a hypothetical infectious disease. In these graphs, three ($M\!=\!3$) types of nodes -- namely adults (type 1), heath-care workers (HCW, type 2) and children (type 3) -- interact whithin groups representing households, workplaces, schools and hospitals. In addition, friendship bonds between children are modeled using a nontrivial motif \mbox{(shown in \fref{fig:allard_fig_4} in \ref{sec:num_sim_details})}, and directed edges from adults and children to HCW are added to account for the susceptibility of HCW to get infected by people seeking care in hospitals \cite{Bansal06_PLoSMed}. The disease spreads from infectious nodes to their neighbours with probability $T$ called the \textit{transmissibility} \cite{Newman02_PhysRevE}. Further details of these graphs and of the associated numerical simulations are relegated to \ref{sec:num_sim_details}. It should be appreciated that these graphs contain a wide range of properties found in real complex networks such as clustering of several orders (e.g., arbitrary motifs, heterogeneous Erd\H{o}s-R\'enyi cliques), (dis)assortative mixing, degree-degree correlation and directed edges.

\begin{figure}[tb]
 \centering
 \includegraphics[width = 0.45\textwidth]{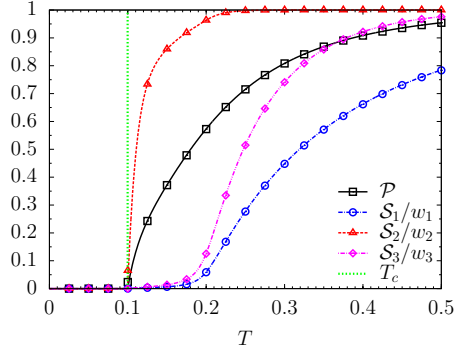}
  \caption{\label{fig:allard_fig_2}(colour online) Bifurcation diagram of the probability to reach the giant component $\mathcal{P}$ and the fraction of nodes of type $i$ therein $\mathcal{S}_i/w_i$ as a function of the occupation probability of edges (or transmissibility) $T$. Types 1, 2 and 3 correspond respectively to adults, HCW and children. Lines represent the theoretical predictions of our formalism [\eref{eq:gc_selfconsistent}--\eref{eq:Si}] while symbols have been obtained through numerical simulations (over $10^5$ simulations on graphs of at least $1.2\times10^5$ nodes for each symbol, see \ref{sec:num_sim_details} for further details). The percolation threshold $T_c \simeq 0.1$ has been obtained by finding the value of $T$ for which the largest eigenvalue of $\mathbf{B}$ equals 1 [see \eref{eq:phase_transition2}].}
\end{figure}

\Fref{fig:allard_fig_2} shows the typical bifurcation diagram of the giant-component-related quantities $\mathcal{P}$ and $\{\mathcal{S}_i\}$. Apart from the excellent agreement between the results of the numerical simulations and the predictions of our formalism, this figure illustrates how the multitype approach can highlight the behavioral differences between different populations --- identified by their own node type --- within a same graph ensemble. In this specific case, the HCW population has purposely been put in the situation where each HCW has more incoming edges than outgoing edges with adults and children. Also, the average degree inside the Erd\H{o}s-R\'enyi cliques corresponding to hospitals (300 nodes connected to one another with probability 0.05) is greater than 1 for $T$ greater than $T^{\prime}\equiv[0.05\times299]^{-1} \simeq 0.067$. This implies that these cliques are increasingly likely to have percolated (i.e., to have a spanning cluster) for $T>T^{\prime}$. Qualitatively, once an outbreak reaches the HCW population, it is likely to stay mostly confined in it and to infect a large proportion of it. Only when $T$ becomes sufficiently large does the outbreak invade other part of the population (schools, workplaces and friendship circles). These insights are corroborated by \fref{fig:allard_fig_2}. It also shows that although the HCW population accounts for only 5\% of the total population, it drives the percolation process by pulling down its threshold to $T_c\simeq0.1$; the other node types only significantly join (i.e., $\mathcal{S}_i/w_i>0.01$) the giant component at $T\simeq0.14$ and $T\simeq0.16$, respectively.

\begin{figure*}[tb]
 \centering
 \subfigure[$\ T<T^{\prime}$]    {\label{fig:allard_fig_3a} \includegraphics[width = 0.45\textwidth]{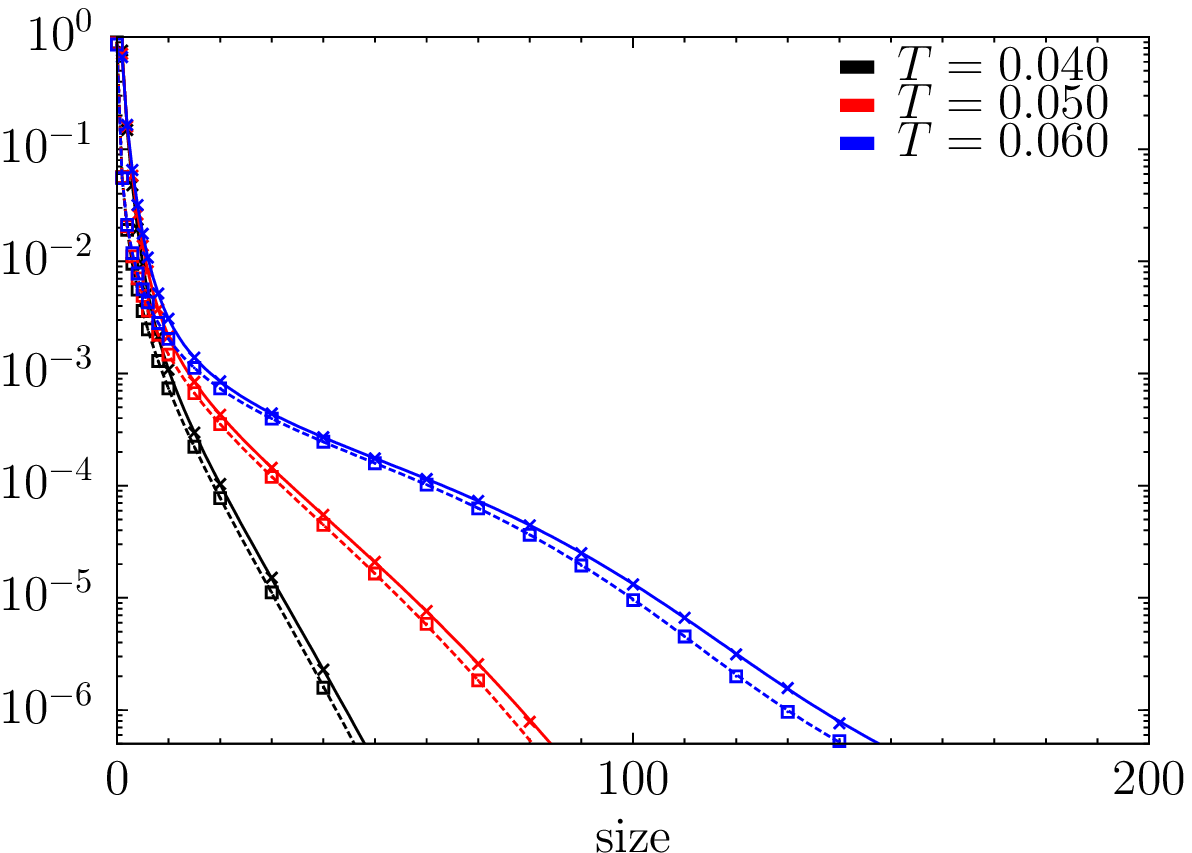}}
 \subfigure[$\ T^{\prime}<T<T_c$]{\label{fig:allard_fig_3b} \includegraphics[width = 0.45\textwidth]{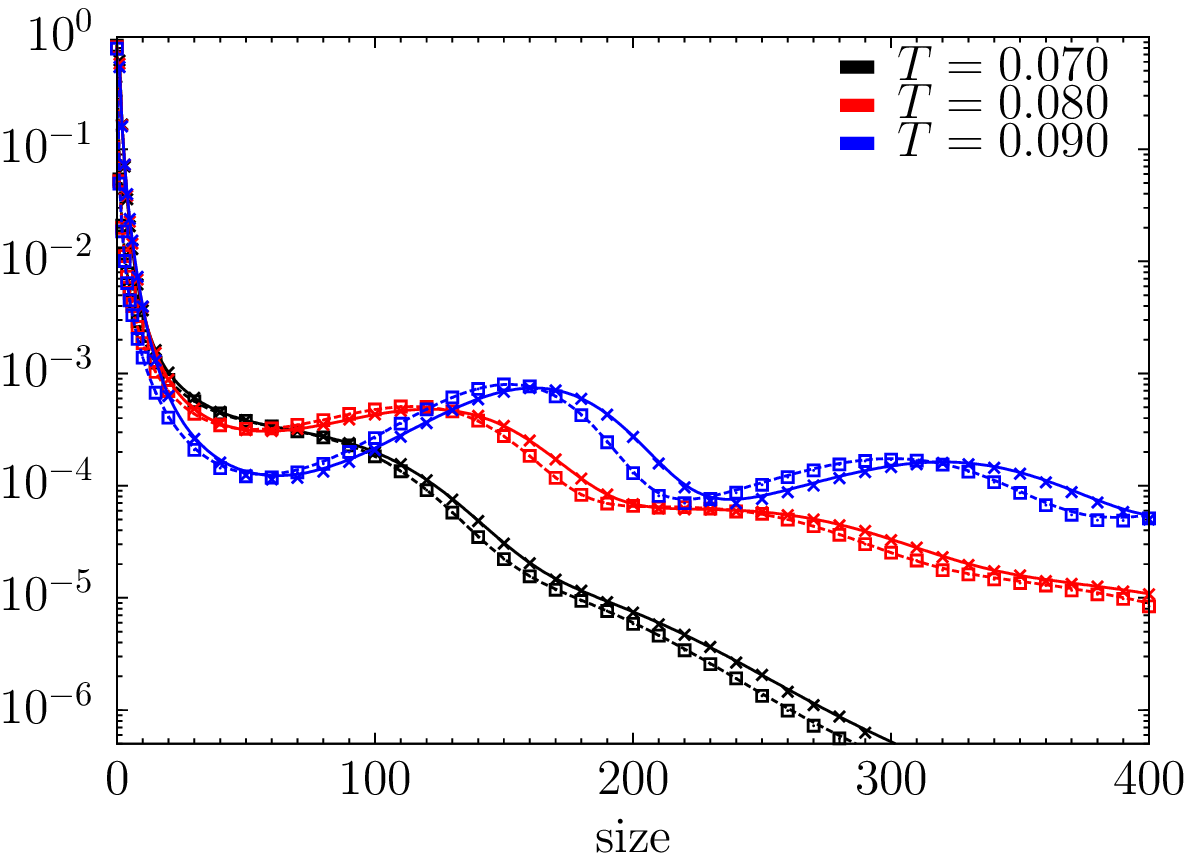}}
 \subfigure[$\ T>T_c$]           {\label{fig:allard_fig_3c} \includegraphics[width = 0.45\textwidth]{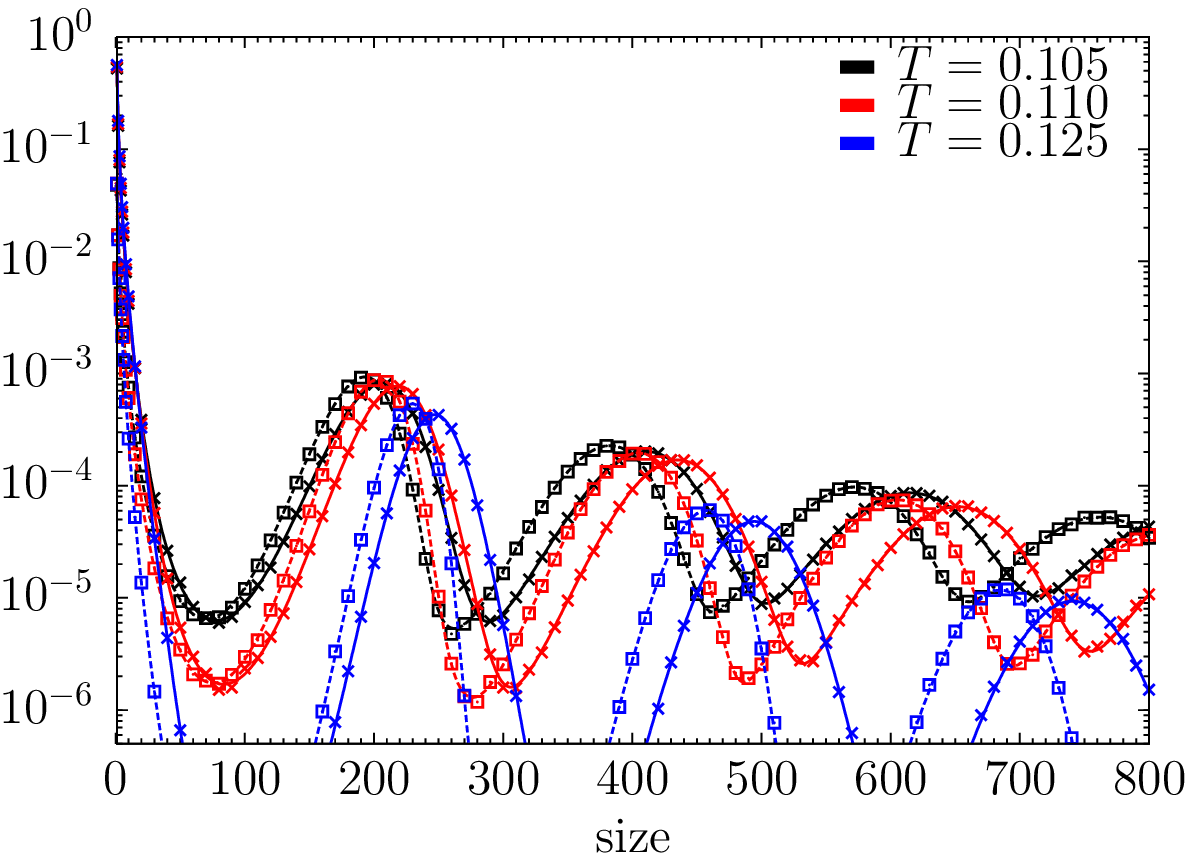}}
  \caption{\label{fig:allard_fig_3}(colour online) Distribution of the number of nodes in small components for various values of the transmissibility $T$ (one colour per value). Continuous and dashed curves represent the total number of nodes and the number of type-2 nodes (HCW) in the small components, respectively. Lines were obtained by solving \eref{eq:sc_translation_independence}--\eref{eq:small_components_size_dist} and symbols were obtained through numerical simulations (over $10^8$ simulations on graphs of at least $4.8\times10^5$ nodes for each symbol, see \ref{sec:num_sim_details} for further details).}
\end{figure*}

\Fref{fig:allard_fig_3} shows the distribution of the total number of nodes in small components for various values of $T$. To support our claim that outbreaks are mostly confined within the HCW populations, \fref{fig:allard_fig_3} also displays the distribution of the number of nodes of type 2 in the small components. The small shift between the two curves is due to adults and children being infected mostly in households. Again, we conclude in an excellent agreement between both the numerical simulations and theoretical predictions of the formalism obtained by solving \eref{eq:sc_translation_independence}--\eref{eq:small_components_size_dist}.

Interestingly, \fref{fig:allard_fig_3a}--\fref{fig:allard_fig_3c} give evidence of what one may call the ``local percolation'' of the hospital cliques as $T$ increases. For $T<T^{\prime}$, the size distribution falls rapidly and monotonously as expected for generic CM graphs \cite{Newman01_PhysRevE,Newman07_PhysRevE}. For $T^{\prime}<T<T_c$, however, the shape of the distribution changes as local maxima appear. These are due to the growing spanning cluster in the hospital cliques. For $T>T_c$, most of the HCW population is part of the giant component, and the spanning cluster is more and more likely to cover the entire clique as $T$ increases. The HCW nodes that are not part of the giant component are therefore likely to be part of very large small components composed of one or more ``locally percolated'' cliques. This is confirmed by the multiple maxima seen on \fref{fig:allard_fig_3b}--\ref{fig:allard_fig_3c}.
%
%
%
%
%
\section{Special cases, generalization and applications} \label{sec:special_cases}
%
We now demonstrate our claims that our formalism encompasses many percolation models on random graphs published in the litterature. We also succinctly outline a possible generalization and some straightforward applications of our model.
%
%
%
\subsection{Multitype random graphs} \label{sec:sc_multitype}
%
Our formalism naturally falls back on the model introduced in \cite{Allard09_PhysRevE} describing the heterogeneous bond percolation on multitype random graphs. In this class of graphs, there are $M$ types of nodes, and a $i \rightarrow j$ edge is occupied with probability $T_{ij}$. Type-$i$ nodes occupy a fraction $w_i$ of the graph, and a type-$i$ node is connected to $\tilde{k}_j$ type-$j$ nodes (for each $j \in [1,M]$) with probability $\tilde{P}_i(\tilde{k}_1,\tilde{k}_2,\ldots,\tilde{k}_M)$.

Our formalism reproduces this model by using one group type for each possible (unordered) type of $i \rightarrow j$ edge. To each of the $\Lambda=M(M+1)/2$ group types are associated the functions
\begin{eqnarray*}
  \theta_{i\nu}(\bm{x}) & = [1 + (x_{\nu j} - 1)T_{ij}] \\
  \theta_{j\nu}(\bm{x}) & = [1 + (x_{\nu i} - 1)T_{ji}]
\end{eqnarray*}
depending whether the edge is considered in the $i \rightarrow j$ or in the $j \rightarrow i$ direction. Along with these functions, $P_i(\bm{k})$ can therefore reproduce the degree distribution $\tilde{P}_i(\tilde{k}_1,\tilde{k}_2,\ldots,\tilde{k}_M)$.

As shown in \cite{Allard09_PhysRevE}, multitype random graphs naturally encompasses multipartite graphs, as well as the undirected random graphs introduced in \cite{Newman01_PhysRevE,Newman02_PhysRevE,Newman03b_PhysRevE}. By assigning nodes with a given degree to a same node type, our formalism can also reproduce degree-degree correlation as in \cite{Newman02_PhysRevLett}.
%
%
%
\subsection{Clustered random graphs}
%
Being a multitype generalization of the \textit{highly clustered random graphs} introduced in \cite{Newman03b_PhysRevE}, our model simplifies to the latter in a straightforward manner with $M=\Lambda=1$ and all groups being Erd\H{o}s-R\'enyi cliques. For $\Lambda=1$, the groups to which any given node belongs to is averaged in \eref{eq:theta_general} so that no correlation whasoever can be taken into account.

When considering only $M=1$ type of nodes but an arbitrary number of uniquely configured groups [$R_\nu(\bm{n})=1$ for all $\nu$], we retrieve \textit{random graphs containing arbitrary distributions of subgraphs} as introduced in \cite{Karrer10_PhysRevE}. The unweighted average \eref{eq:unweighted_average} plays an analogous function as their \textit{role} distribution, with correlation being taken into account by using node types. It is then straightforward to conclude that our formalism also encompasses the \textit{edge-triangle} model introduced in \cite{Miller09_PhysRevE,Newman09_PhysRevLett} and the \textit{strong ties} model proposed by \cite{Shi07_PhysicaA}.

The \textit{$\gamma$-theory} model \cite{Gleeson09_PhysRevE} can be recovered by considering only $M=1$ type of nodes, and by allowing nodes to belong to only one group of size larger than two (Erd\H{o}s-R\'enyi cliques) but to belong to an arbitrary number of group of size two (external edges). Also, the \textit{random hypergraphs} introduced in \cite{Ghoshal09_PhysRevE} can be reproduced by our formalism by considering $M=3$ types of nodes and $\Lambda=1$ type of groups which are triangles composed of one node of each type.

Finally, a class of formalism \cite{Serrano06_PhysRevLett,Zlatic12_EPL,Serrano06b_PhysRevE} uses the \textit{multiplicity} of edges --- the number of triangles to which an edge participates --- to derive an effective branching process and solve the percolation on clustered graphs using PGFs. Although this approach tackles percolation from a different perspective, its predictions (\textit{i.e.}, the percolation threshold and the size of the giant component) can be reproduced with our model by using fully connected motifs of size $m+2$ to account for links of multiplicity $m$ and by appropriately using node and group types to account for the correlations that this class of models incorporates.
%
%
%
\subsection{Directed random graphs}
%
Our formalism as presented in this paper can only model directed edges between \emph{different} node types. To describe directed edges among a same node type such as in \cite{Newman01_PhysRevE,Meyers06_JTheorBiol}, we would need to subdivide the group type corresponding to directed edges into an incoming part and an outgoing part (e.g., $\nu \rightarrow \nu_\mathrm{in}, \nu_\mathrm{out}$), and match the complementary parts to form groups in the unclustered multitype bipartite graph. In other words, each group is linked to an incoming and an outgoing stub. The mathematical formalism introduced in \sref{sec:math_form} remains valid except that we would need to explicitly consider the fact that nodes are reached by their incoming edges and are left by their outgoing edges when writing down the equations (see \cite{Newman01_PhysRevE,Meyers06_JTheorBiol} for detailed examples). This adjustment is nevertheless straightforward and does not affect the generality of our approach.
%
%
%
\subsection{Interdependent or coupled networks}
%
The use of node and group types naturally permits our formalism to be used in the study of interdependent or coupled networks. In interdependent -- or interacting -- networks, node types could for instance be used to distinguish the elements of two or more interacting networks \cite{Leicht09_arXiv,Buldyrev10_Nature,Huang11_PhysRevE}. Different group types would then allow to specify precisely the (nontrivial) interactions within and across the networks. In the case of coupled, or overlayed, networks \cite{Marceau11_PhysRevE,Funk10_PhysRevE} elements in a single population interact in different ways which is modelled using different edge types. This again can be easily achieved with our formalism by defining multiple group types, one for each level of interaction. Again, the generality of our approach gives us access to a wide variety of complex patterns of interactions in a very detailed fashion.
%
%
%
\subsection{\label{sec:weak_and_strong}Weak and strong clustering regimes}
%
The existence of two regimes of clustering, \textit{weak} and \textit{strong}, has been put forward in \cite{Serrano06_PhysRevLett,Serrano06a_PhysRevE,Serrano06b_PhysRevE} with the conclusion that these two regimes have opposite effect on the bond percolation threshold. In the weak regime, edges have a multiplicity of either 0 or 1 (single edges or disjoint triangles), and the percolation threshold is higher than for equivalent unclustered graphs. In the strong regime, edges may contribute to more than one triangle, and it is argued that the percolation threshold is then lower than for equivalent unclustered graphs.

Contrariwise, the analysis done in \cite{Miller09_PhysRevE,Gleeson09_PhysRevE} strongly suggests that clustering always increases the percolation threshold and that the observed lower percolation threshold in the strong regime is due to assortative mixing instead. Hence, according to theses results, there should be no weak and strong clustering regimes. The use of node and group (or edge) types in our model can generate clustered and unclustered graphs with the same correlations (or mixing patterns). It is therefore possible to investigate --- both numerically and analytically --- the effect of clustering alone on the percolation threshold, shedding some light on this contradiction while extending the analysis and the conclusions of \cite{Miller09_PhysRevE,Gleeson09_PhysRevE}. This will be addressed in a future publication.
%
%
%
%
%
\section{Conclusion} \label{sec:conclusion}
%
We have presented a generalization of the Configuration Model allowing for the inclusion of several nontrivial mixing patterns and clustering. On the one hand, the use of node and group types permits to explicitly prescribe how nodes are connected to one another, hence reproducing (dis-)assortative mixing, and indirectly degree-degree correlation. On the other hand, the use of a one-mode projection can generate a wide range of nontrivial clustered structures through quenched or annealed motifs. Besides the modeling of mixing patterns, the multitype approach permits to identify nodes. This allows to highlight unusual behaviors or susceptibility of sub-population of nodes, as well as to simulate targetted intervention such as attacks, failures, vaccination or quarantine. We have also demonstrated that our formalism encompasses several models published to this day, and we have outlined potential applications.

Bridging the gap between empirical network datasets and theoretical models is surely one the principal tenets of network theory. Since extracting the effective clustered backbone (\textit{i.e.}, motifs) of real networks is still an open problem, our approach can only offer a partial answer. However, it provides a comprehensive synthesis of the many variants of the CM published to date, and it extends considerably the structural complexity of graphs that can be handled theoretically. In these regards, the versatility and generality of the present framework could prove useful even beyond the strict confines of bond percolation on complex graphs.
%
%
%
%
%
\ack
%
The authors would like to thank Vig\'e Lebrun for fruitful and sobering discussions. We are also grateful to an anonymous referee for pointing out a number of relevant publications. This work has been supported by the CIHR, the NSERC and the FRQ-NT.
%
%
%
%
%
\appendix
\section{General method to compute $Q_{i\nu}(\bm{l}|\bm{n})$ for arbitrary multitype motifs} \label{sec:general_method}
%
We present a systematic way to compute the outcome of bond percolation, $Q_{i\nu}(\bm{l}|\bm{n})$, on any arbitrary multitype motifs where edges are simple and can be directed or not.

Let us first consider a multitype generalization of Erd\H{o}s-R\'enyi random graphs. These are composed of $\bm{n}$ nodes, and a directed edge exists from a type-$i$ node to a type-$j$ node with probability $p_{ij}$. Edges exist independently of one another. Note that the symmetric case $p_{ij}=p_{ji}$ is statistically equivalent to undirected edges. It has been shown \cite{Allard12_EPL} that $Q_{i\nu}(\bm{l}|\bm{n})$ can be obtained by iterating 
\begin{eqnarray}
 Q_{i\nu}(\bm{l}|\bm{n}) & = Q_{i\nu}(\bm{l}|\bm{l}) \prod_{rs} {n_r\!-\!\delta_{ir} \choose l_r\!-\!\delta_{ir}} (1\!-\!p_{rs})^{l_r(n_s-l_s)} \label{eq:iterative_erdos_renyi_1} \\
\fl \mbox{and} \nonumber \\
 Q_{i\nu}(\bm{l}|\bm{l}) & = 1 - \sum_{\bm{m}<\bm{l}} Q_{i\nu}(\bm{m}|\bm{l}) \label{eq:iterative_erdos_renyi_2}
\end{eqnarray}
from the initial condition $Q_{i\nu}(\bm{\delta_i}|\bm{\delta_i}) = 1$ with $\bm{\delta_i} \equiv (\delta_{i1},\ldots,\delta_{iM})$. In essence, knowing the probability of finding a component of size $\bm{l}$ from a node of type $i$ in a graph of size $\bm{l}$, \eref{eq:iterative_erdos_renyi_1} computes the probability of finding a sub-component of size $\bm{l}$ but in a graph of size $\bm{n}$ ($>\bm{l}$). This allows to compute every coefficients of the distribution $Q_{i\nu}(\bm{l}|\bm{n})$ except for the last one, the one corresponding to the case where the whole graph is reachable, which is obtained using \eref{eq:iterative_erdos_renyi_2}.

Let $\mathcal{G}$ be a multitype motif composed of $\bm{n}$ nodes with an arbitrary configuration of edges. The associated distribution $Q_{i\nu}(\bm{l}|\bm{n})$ can then be computed by following these simple steps:
\begin{enumerate}
 \item Consider an equivalent multitype Erd\H{o}s-R\'enyi graph $\mathcal{G}'$ of size $n'=\sum_j n_j$ in which each node belongs to its own unique type (i.e., $n'_{j'}=1$ for all $j' \in \{1,\ldots,n'\}$). Note $p'_{i'j'}$ the probability for a directed edge to exist from the type-$i'$ node to the type-$j'$ one.
 \item Compute $Q'_{i'\nu}(\bm{l'}|\bm{n'})$ for $\mathcal{G}'$ with \eref{eq:iterative_erdos_renyi_1}--\eref{eq:iterative_erdos_renyi_2}. Without any loss of generality suppose that the initial node from which the graph is probed is of \mbox{type 1} [i.e., \eref{eq:iterative_erdos_renyi_1}--\eref{eq:iterative_erdos_renyi_2} need to be solved only once].
 \item From $Q'_{i'\nu}(\bm{l'}|\bm{n'})$, derive the intermediate distribution $Q_{i\nu}^{(j)}(\bm{l}|\bm{n})$ of the number of nodes of each type that are accessible from the $j$-th type-$i$ node. This is achieved by replacing the \emph{artificial} node types in $\mathcal{G}'$ by the \emph{actual} node types in $\mathcal{G}$, and by setting the values of $p'_{i'j'}$ according to the configuration of the edges in $\mathcal{G}$, which can include type-dependent probabilities of existence/occupation of edges.
 \item Obtain $Q_{i\nu}(\bm{l}|\bm{n})$ by computing the unweighted average of the $n_i$ distributions $Q_{i\nu}^{(j)}(\bm{l}|\bm{n})$
 \begin{equation} \label{eq:unweighted_average}
  Q_{i\nu}(\bm{l}|\bm{n}) = \frac{1}{n_i}\sum_{j} Q_{i\nu}^{(j)}(\bm{l}|\bm{n}) \ .
 \end{equation}
\end{enumerate}
A noteworthy point is that the distribution $Q'_{i\nu}(\bm{l'}|\bm{n'})$ computed for a generic graph of size $n'$ can generate every multitype motif of size smaller than $n'$ by appropriate choices of $p'_{i'j'}$. An explicit example of such a calculation is given in \cite{Allard12_EPL}.
%
%
%
%
%
\section{Numerical simulations} \label{sec:num_sim_details}
%
\begin{table}
 \caption{\label{tab:R}Distribution $R_{\nu}(\bm{n})$ used for the simulations in \sref{sec:validation} with $M=3$ and $\Lambda=7$.}
 \begin{indented}
  \item[] \begin{tabular}{ c c c}
    \br
    Group type & Composition              & Probability       \\
               & $\bm{n} = (n_1,n_2,n_3)$ & $R_{\nu}(\bm{n})$ \\
    \mr
    \multirow{12}{*}{Households} & $(2,0,0)$ & 0.0810 \\
                                 & $(1,1,0)$ & 0.0180 \\
                                 & $(0,2,0)$ & 0.0010 \\
                                 & $(2,0,1)$ & 0.1215 \\
                                 & $(1,1,1)$ & 0.0270 \\
                                 & $(0,2,1)$ & 0.0015 \\
                                 & $(2,0,2)$ & 0.3240 \\
                                 & $(1,1,2)$ & 0.0720 \\
                                 & $(0,2,2)$ & 0.0040 \\
                                 & $(2,0,3)$ & 0.2835 \\
                                 & $(1,1,3)$ & 0.0630 \\
                                 & $(0,2,3)$ & 0.0035 \\
    \mr
    \multirow{3}{*}{Schools} & $(5,0,50)$   & 0.2500 \\
                             & $(10,0,100)$ & 0.5000 \\
                             & $(15,0,150)$ & 0.2500 \\
    \mr
    \multirow{5}{*}{Workplaces} & $(10,0,0)$ & 0.1000 \\
                                & $(20,0,0)$ & 0.2500 \\
                                & $(30,0,0)$ & 0.3000 \\
                                & $(40,0,0)$ & 0.2500 \\
                                & $(50,0,0)$ & 0.1000 \\
    \mr
    Hospitals & (0,300,0) & 1.0000 \\
    \mr
    Friendships & (0,0,5) & 1.0000 \\
    \mr
    Directed edges (1 $\rightarrow$ 2) & (1,1,0) & 1.0000 \\
    \mr
    Directed edges (3 $\rightarrow$ 2) & (0,1,1) & 1.0000 \\
    \br
  \end{tabular}
 \end{indented}
\end{table}

Details of the graphs used in \sref{sec:validation} and of the numerical simulations performed to validate our formalism are presented.
%
%
%
\subsection{Urban networks}
\begin{figure}[t]
 \centering
 \includegraphics[width = 0.20\textwidth]{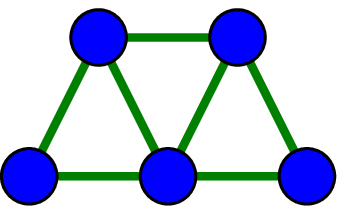}
 \caption{\label{fig:allard_fig_4}Motif used to model friendship bonds between children in the \textit{urban network} used in \sref{sec:validation}.}
\end{figure}

%
The graphs generated in \sref{sec:validation} were inspired by the \textit{urban networks} used in \cite{Meyers05_JTheorBiol,Bansal06_PLoSMed} in which individuals are connected to one another because of their common membership to a social group (e.g., households, schools, workplaces, hospitals, friendship circles). In this case, the population is divided into three categories -- identified by node types -- namely adults (type 1), health-care workers (HCW, \mbox{type 2)} and children (type 3) with $\{w_i\} = \{0.45,0.05, 0.50\}$. Every node belongs to one household, every HCW belongs to one hospital, every child belongs to one school, 1/9 of adults belong to one school (teachers, janitors, etc.) and the remaining 8/9 belong to one workplace. Also every child belongs to one group of friends (see \fref{fig:allard_fig_4}), and adults and children are connected at most to two randomly chosen HCW via a directed edge.

\Tref{tab:R} explicits the group composition distribution $R_{\nu}(\bm{n})$ used to generate the urban networks. Except for friendship circles, the connections between individuals within groups are modeled with multitype Erd\H{o}s-R\'enyi graphs with different probabilities of edge existence. In households, every possible edge exists except for the directed edges from HCW to adults, HCW and children that exist with probability 0.2, 0.2 and 0.1, respectively. In schools and workplaces, edges exist with probability 0.01 and they exist with probability 0.05 in hospitals. The use of relatively large cliques with such low probabilities of existence of edges allows to model redundancy in the neighbourhood of nodes while keeping a relatively low clustering. Finally, directed edges from adults and children to HCW exist with probability 0.5.

These graphs can be generated in a fairly straightforward manner. For a given group type $\nu$, we first generate a sequence of groups whose composition is prescribed by $R_{\nu}(\bm{n})$. We then generate, according to $P_i(\bm{k})$, a list of nodes in which a node belonging to $k_\nu$ type-$\nu$ groups appears $k_\nu$ times. We finally randomly assign these nodes to the groups, and create edges between nodes that are members of a same group according to the probabilities given in the last paragraph.
%
%
%
\subsection{Percolation simulations}
%

Graphs that were used to obtain the results shown in \sref{sec:validation} were composed of at least 1.2$\times 10^5$ nodes. For $T$ around $T_c$, larger graphs (up to 9.6$\times 10^6$ nodes) have been generated to faciliate the distinction between small components, which are intensive, from the giant component, which is extensive. At least $10^3$ ($10^6$) graphs were generated for each value of $T$ used in \fref{fig:allard_fig_2} (\fref{fig:allard_fig_3}).

For each generated graph, 100 percolation simulations were performed. These consist in randomly choosing a starting node and then following every possible edges leaving this node -- and the subsequently encountered nodes -- with probability $T$ until no new node can be reached. The component size is then simply the number of nodes that have been reached. While it would have been straightforward to use a type-specific probability $T$ (see \sref{sec:sc_multitype}), we have used a single value to lighten the presentation of the results.
%
%
%
%
%
\section*{References}
%

%
%
%
\end{document}